\documentclass[aps,prl,twocolumn,groupedaddress,nofootinbib]{revtex4-2}

\usepackage{graphicx}
\usepackage{physics}
\usepackage{xcolor}
\usepackage{verbatim}

\usepackage{textpos} 
 
\begin{document}

\begin{textblock}{9}(-2,-0.7) \textit{To be published in Phys. Rev. Lett. (first version submitted 12/27/2023) }  \end{textblock}

\vspace{-12pt}


\title{ \vspace{-0.5cm}  
The persistence of spin coherence in a crystalline environment}

\author{Gerald Curran III, Zachary Rex, Casper X. Wilson, Luke J. Weaver, Ivan Biaggio}
\affiliation{Department of Physics, Lehigh University, Bethlehem, PA 18018, USA}
 
\date{\today}

\begin{abstract}
We analyze quantum interference in the triplet-exciton pair generated by singlet exciton fission in a molecular crystal, and introduce transport-induced dephasing (TID) as a key effect that can suppress  the expected fluorescence quantum beats when the triplet-exciton wavefunction can localize on inequivalent sites. TID depends on the triplet-exciton hopping rate between inequivalent sites and on the energy-shifts among the stationary states of the entangled triplet pair in different spatial configurations. The theoretical model is confirmed by experiments in rubrene single crystals, where triplet pairs remain entangled for more than 50 ns but quantum beats  are  suppressed by TID within a few nanoseconds when the magnetic field is misaligned by just a few degrees from specific symmetric directions. Our experiments deliver  the  zero-field parameters for the rubrene molecule in its orthorhombic lattice and  information on triplet-exciton transport, in particular the triplet-exciton hopping rate between inequivalent sites, which we evaluate to be of the order of 300 ps in rubrene.
\end{abstract}


\maketitle

Photoexcitation in  organic molecular crystals may result in singlet excitons that undergo a fission process to create a pair of mobile, spin-entangled triplet excitons with overall zero spin. 

When the  exciton in a triplet pair has  near half the singlet energy,   the  emissive singlet state can be recreated by a geminate triplet fusion process. Time-modulation of the resulting fluorescence (quantum beats) then becomes possible for the spin-entangled triplet-pair  because of quantum interference upon fusion, as seen   in  single crystals of tetracene \cite{Chabr81, Funfschilling85,Funfschilling85b} and rubrene \cite{Wolf18}. 

The entangled triplet-exciton pair,   an analog to entangled photon pairs  \cite{Pan12},  is a  unique condensed-matter candidate for quantum-information studies  \cite{Bardeen19,Smyser20}.  However, to date no investigations have focused on the coherence of the spin wavefunction as   the triplet excitons separate from each other and travel independently in the crystal. In this work, we use quantum interference between the excitons to evaluate the effects of transport.

Instead of  controlling the path traveled by the  entangled triplet excitons,  we exploit the fact that   fusion events observed at different times  are affected by the different distances (that can reach up to hundreds of nanometers in our experiments) traveled by the  excitons   before their re-encounter. The  time-dependence of the observed quantum beats then provides information on the triplet-pair and how it is affected by exciton transport.

We find that, in contrast to early investigations in various acenes  \cite{Sternlicht61, Merrifield69, Johnson70, Suna70}, and in particular   in tetracene, where triplet excitons average over inequivalent sites \cite{Yarmus72} and always lead to  quantum beats  \cite{Chabr81, Funfschilling85,Funfschilling85b},  triplet excitons in rubrene  localize on  inequivalent sites, similar to what was observed in a solid solution of naphtalene \cite{Hutchison61}. Triplet exciton hopping between inequivalent sites then leads to a corresponding change in the Hamiltonian of the triplet pair, causing  a transport-induced dephasing (TID) of its spin wavefunction  that can destroy the quantum beats. This finding will   require  a reevaluation of the assumptions used in other systems, and it also provides a new way to assess the extent of the triplet-exciton wavefunction or hopping times between inequivalent sites, and  characterize triplet-exciton transport.

\begin{figure}
\includegraphics[width=\columnwidth]{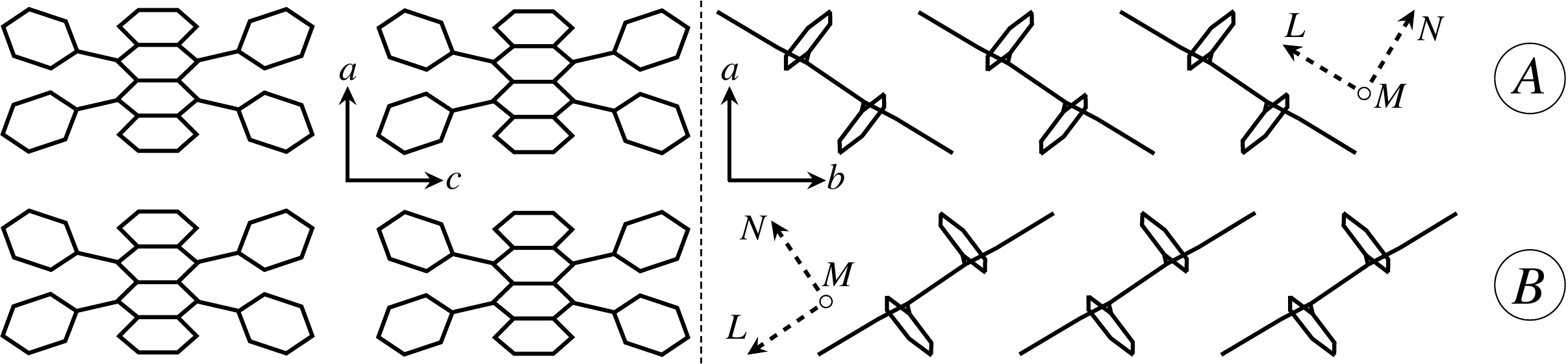}
\caption{Adjacent $\pi$-stacked molecular columns (represented by their carbon-carbon bonds) in the  lattice of orthorhombic rubrene \cite{jurchescu06}, in  $b$-axis (left) and  $c$-axis (right) projections, with the   $a$, $b$, $c$ crystal axes used in this work. The unit cell of rubrene contains two differently oriented inequivalent molecules (labeled $A$ and $B$). Their  $L$, $M$, and $N$  coordinate system are drawn with dashed lines.   The angle between the  $L$-axis  and the $b$-axis is $\pm 31.0^\circ$ (from the room-temperature X-ray data in Ref.~\onlinecite{jurchescu06}). The molecules in the $b$-axis projection are  intercalated in the $b$ direction, while those in the $c$-axis  projection are all at the same $c$-coordinate.  Triplet-exciton diffusion is most favored along the $b$-axis and least favored along the $c$-axis \cite{Wolf21}. \label{RubreneXstalViews}}
\end{figure}

Since we are presenting experimental data in orthorhombic rubrene, we perform all our calculations for its crystal structure   \cite{jurchescu06}, shown in  Fig.~\ref{RubreneXstalViews}.  
We start with the standard  dipole-dipole interaction Hamiltonian for two electrons in a triplet state  that is  written in terms of  the  spin vector and the $D$ and $E$ ``zero field splitting''' parameters that specify the stationary state energies for an individual triplet exciton, and are generally used in electron paramagnetic resonance experiments \cite{Yarmus72,Swenberg73}. We then derive the Hamiltonian for each inequivalent site by doing a coordinate transformation  \cite{Tapping16,Piland16} to the crystal coordinate system where $x=a, y=b, z=c$ (rotation by $\pm 31.0^\circ$ around the $c$-axis, see Fig.~\ref{RubreneXstalViews}), and then adding the Zeeman contribution. As spin basis states we use either the standard zero-field states $\ket{x}$, $\ket{y}$, and  $\ket{z}$  that have a  zero  $x$-, $y$-, or $z$-component of the spin, respectively (similar to  the $p$-states of the hydrogen atom), or the high-field states   $\ket{1}, \ket{0}, \ket{-1}$ defined in terms of the  component of the  spin along the direction of an applied magnetic field. 
We then construct the product states and Hamiltonian for the triplet-pair. We find that it is a good initial approximation to use the tetracene molecular parameters  $D = 0.052$ cm$^{-1}$ and $E = - 0.0052$  cm$^{-1}$  \cite{Yarmus72} for the tetracene backbone of rubrene,  and use $g=2$ for the gyromagnetic factor. We also neglect the magnetic dipole-dipole interaction between  triplet excitons because it decays with the cube of the distance \cite{Tapping16}, and  triplet excitons  separate rapidly  and spend most of the time away from each other \cite{Wolf21}.
We will show below that the predictions from this model are in  good agreement with our observations in rubrene.

\begin{figure}
\includegraphics[width=\columnwidth]{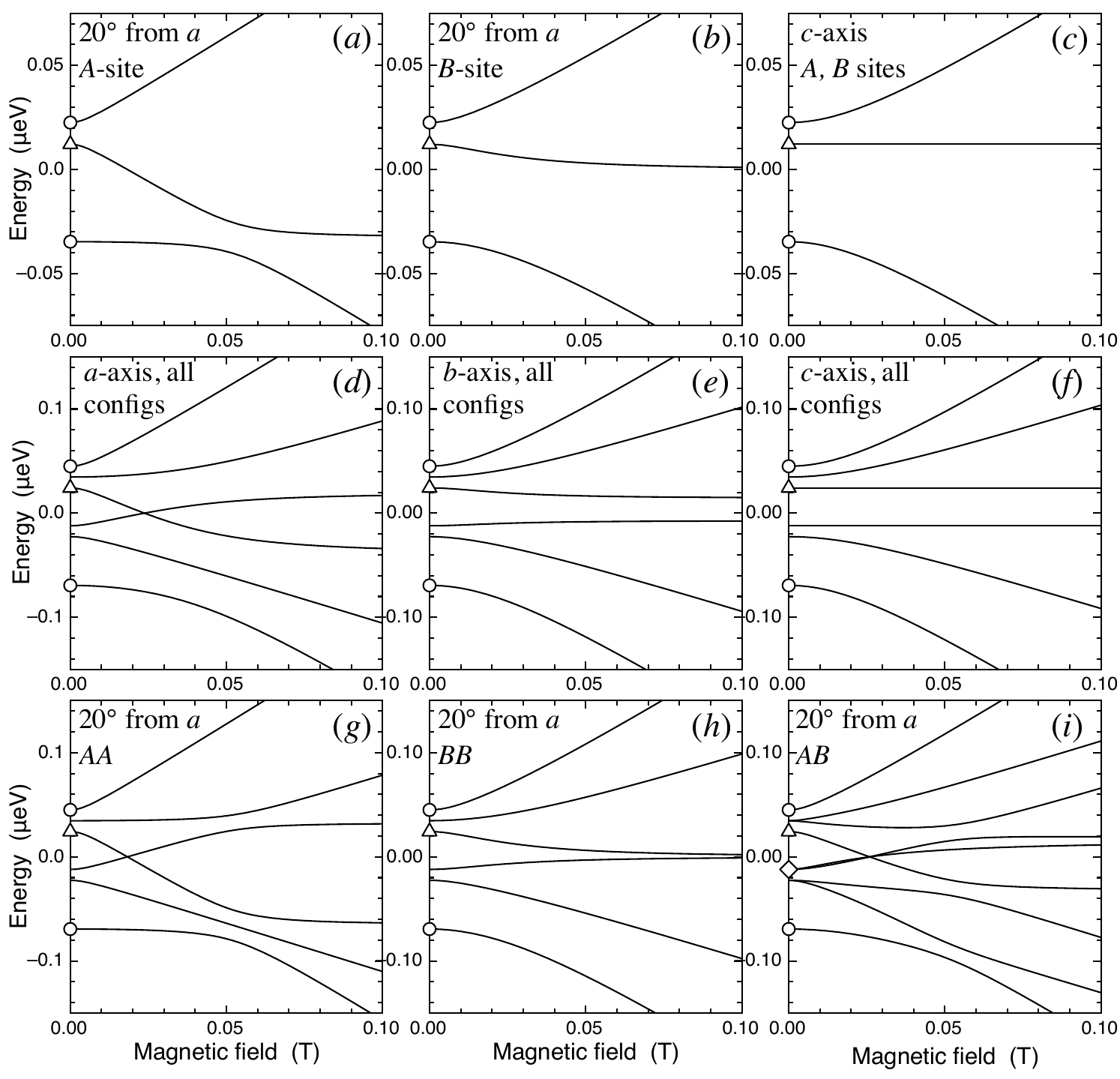}
\caption{Magnetic field dependence of the stationary state energies of a triplet exciton and a triplet-exciton pair  in rubrene for different spatial configurations  (labeled with $AA$, $BB$, $AB$), calculated using tetracene molecular parameters. Energies for  individual triplet excitons are for magnetic fields at $20^\circ$ from the $a$-axis and along the $c$-axis ($a$-$c$). Energies for triplet-pairs  are for magnetic field along the main axes ($d$-$f$), and for  magnetic field at $20^\circ$ degrees from the $a$ axis in the $ab$-plane ($g$-$i$). The symbols on the vertical axes mark the  stationary states  that contribute to the singlet-state wavefunction of the triplet-pair. States  that are a combination of the $\ket{x}$ and $\ket{y}$ states  in the crystal reference frame are marked with an open circle. The  $\ket{z}$ states are marked wth a triangle. An additional stationary state contributing to the overall singlet state that exists for a triplet pair in  the $AB$ configuration is marked by a diamond symbol in panel ($i$).
\label{EnergiesVsField}}
\end{figure}

A key point here is that the Hamiltonian of the triplet pair   depends in general on its spatial configuration, that is on which of the two inequivalent molecules (top row or bottom row in Fig.~\ref{RubreneXstalViews})   each triplet exciton can be assigned to. We label the three possible spatial configurations of the triplet pair as: $AA$, when both excitons are assigned to molecules rotated by  $+31^\circ$ away from the $b$-axis, $BB$ when both excitons are assigned to molecules rotated by  $-31^\circ$ away from the $b$-axis, or $AB$ when each of the two excitons is assigned to a different molecular family.  

Fig.~\ref{EnergiesVsField}  shows the prediction of the above model for the magnetic field dependence of the energy eigenvalues  of triplet excitons  in different spatial configurations. 

At zero magnetic field, all energy eigenvalues are the same for all spatial configurations but, because of the different orientation of the two inequivalent molecules, the corresponding stationary states belong to different linear combinations of spin basis states. In general, an individual triplet exciton with a given spin  may be in a different linear combination of stationary states depending on which inequivalent site it occupies. Since we are interested in the evolution of triplet pairs entangled in an overall singlet  state, we add symbols to the energy axes of Fig.~\ref{EnergiesVsField} to mark  stationary states that contribute to the zero-spin wavefunction at zero field. Note how the effect of the different orientations of the two inequivalent molecules becomes strikingly visible for the  $AB$ spatial configuration (see  Fig.~\ref{EnergiesVsField}$i$), which has four stationary states contributing to its overall spin-zero wavefunction  instead of just three otherwise.

In the presence of a magnetic field, the situation becomes more complicated, but it simplifies again in the high-field limit, seen in Fig.~\ref{EnergiesVsField}$d$-$i$ when stationary state energies become linearly dependent on the magnetic field magnitude, with slopes given by the spin component along the magnetic field.
 
In general, the entangled triplet-pair state created by singlet fission must have a spin wavefunction $\ket{\psi(t)}$ that is initially the overall singlet state, and  then evolves according to the triplet-pair Hamiltonian. The geminate fusion probability is then modulated by  $|\braket{\psi(0)}{\psi(t)}|^2$~\cite{Chabr81}. This can  lead to fluorescence quantum beats when the spatial configuration of the triplet-pair state does not affect the hamiltonian, or remains constant. From Fig.~\ref{EnergiesVsField}, we would then expect zero-field quantum beats to be modulated by  three frequencies for a constant  $AA$  configuration (three stationary states), or by six frequencies for the $AB$ configuration (four stationary states).

However, when the spatial configuration continuously and stochastically changes as triplet excitons hop between inequivalent sites, the time-evolution of the triplet-pair wavefunction  $\ket{\psi(t)}$ becomes irregular, and accumulated fluctuations in hopping time can then lead to random dephasing of the stationary states making up the triplet-wavefunction. At zero magnetic field, this always  suppresses the quantum beats of a triplet-pair population whenever the hopping time is large enough.  
But in the high-field limit, $\ket{\psi(t)}$ is \cite{Chabr81}
\begin{equation}
\ket{\psi(t)} = \tfrac{1}{\sqrt{3}} \left[ e^{-i E_0 t/\hbar} \ket{ 0, 0} -  e^{-i E_1 t/\hbar}  \left( \ket{1,-1} +  \ket{-1,1} \right) \right] \label{pairwavefunction}
\end{equation}
where    the energies  $E_0$ and $E_1$   correspond to the horizontal asymptotes towards which some of the stationary state energies in Fig.~\ref{EnergiesVsField}$d$-$i$ tend as the magnetic field strength increases. The $E_i$ ($i = 0,1$)  depend in general on the spatial configuration of the triplet pair, but their values are always related by   $E_i^{AA} + E_i^{BB} = 2 E_i^{AB}$.

Whenever $E_i^{AA} \neq E_i^{BB}$, triplet-exciton transport leads to TID and the suppression of  quantum beats. However, these energies are equal  when the  magnetic field is parallel to the symmetry planes of the inequivalent molecules (parallel to the $ac$ or $bc$ planes in rubrene, the case shown in the second row of Fig.~\ref{EnergiesVsField}).  This  eliminates TID  as soon as the high-field limit is reached, and the probability of photon emission through triplet-exciton fusion   can then be easily calculated to be  $|\braket{\psi(0)}{\psi(t)}|^2 = 5/9 [1 + 4/5 \cos( \omega t)]$, where $ \omega = (E_0 - E_1)/\hbar$ is the angular frequency of the quantum beats.

\begin{figure}
\includegraphics[width=\columnwidth]{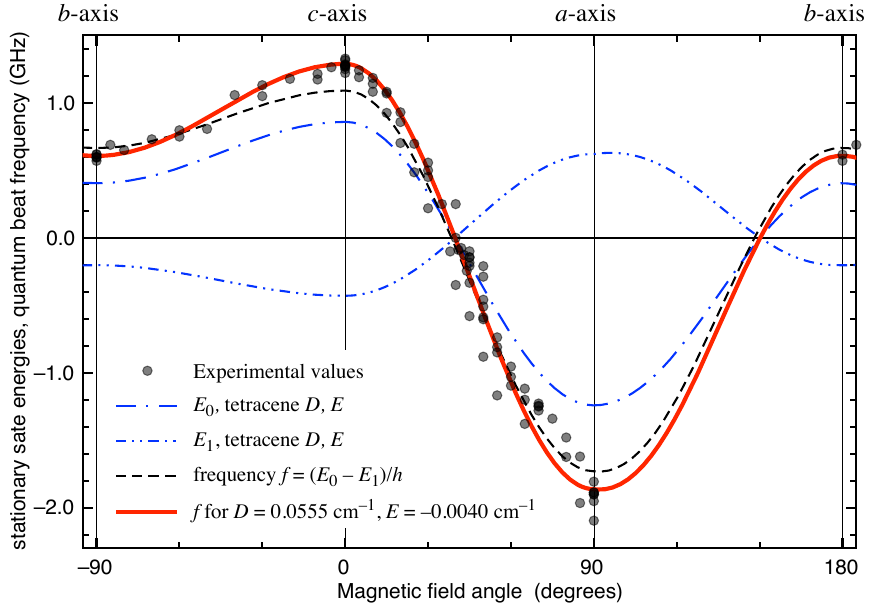}
\caption{High-field quantum beat frequency as a function of the direction of a 0.3 T magnetic field parallel to the main crystallographic planes of rubrene. Experimental results over multiple samples are represented with semi-opaque data points that become darker for  multiple measurements at the same magnetic field orientation. No data is seen in the $ab$-plane because TID leads to destructive quantum interference. Values for $E0$, $E_1$, and $f = (E_0 - E_1)/h$ calculated from  tetracene molecular parameters are given by the dashed lines. The solid red curve is the best fit to the data that determines the  $D$ and $E$ rubrene molecular parameters. \label{FrequencyVsAngle}}
\end{figure}

We confirmed this prediction in rubrene by measuring the magnetic-field-induced quantum beats via time-correlated single photon counting,  using 150~fs, 513~nm pulses obtained  from a Light Conversion PHAROS laser operating at a repetition rate of $200$ kHz.

Fig.~\ref{FrequencyVsAngle} shows the experimental results together with the theoretical predictions for the pseudo-frequency $f = (E_0 - E_1)/h$, which changes sign and becomes negative when $E_1 > E_0$.  The model based on the tetracene molecular parameters (dashed curve) already shows good agreement with the experiment, but this data can now be used to obtain the correct $D$ and $E$ values for the rubrene molecule in its orthorhombic crystal lattice. The experimental values for the beat frequencies for $\vec B \parallel a,b,c$ are $f_a = 1.90 \pm 0.02$ GHz, $f_b = 0.60 \pm 0.05$ GHz,  and $f_c = 1.29 \pm 0.02$ GHz, consistent  with the  theoretical requirement that $f_a = f_b + f_c$. A least-squares fit of the theoretical frequency to these on-axis experimental values and the data in Fig.~\ref{FrequencyVsAngle} delivers  $D = 0.0555$~cm$^{-1}$  and $E = - 0.0040$~cm$^{-1}$ within a confidence interval of $\pm 1\%$.  Since a triplet state does not average between the  inequivalent sites in the rubrene crystal, it does not make sense to define ``crystal'' $D^*$ and $E^*$ values. 

The effect of TID can be evaluated by considering the random walk of the energies $E_0$ and $E_1$ (of  the high-field stationary states $\ket{ 0, 0} $ and $\ket{1,-1} +\ket{-1,1}$)  due to the random hopping of triplet excitons  between inequivalent sites. For a random walk of duration $t$, the total ``dwell times''  spent in each configuration are $t_{AA} = t_{BB} = t/4$ and $t_{AB} = t/2$. But for any given random walk there will be stochastic variations on these dwell times. The time-dependent phase of  the state $\ket{ 0, 0}$ is then given by $\hbar \varphi_0(t) = E_0^{AB}(t/2+\Delta t^{AB}) + E_0^{AA}(t/4+\Delta t^{AA}) + E_0^{BB}(t/4+ \Delta t^{BB})$, and similarly for $\varphi_1(t)$. Since the random walk necessarily goes from one configuration  to the other, any additional time spent in one configuration  implicitely means less time spent in another, and one must have $\Delta t^{AB} + \Delta t^{AA} + \Delta t^{BB}~=~0$. Combining this condition with $\varphi(t) = \varphi_0(t) - \varphi_1(t)$ and  $E_i^{AA} + E_i^{BB} = 2 E_i^{AB}$ leads, after a little algebra, to  $|\braket{\psi(0)}{\psi(t)}|^2$ being modulated by $\cos \varphi(t)$, with
$\varphi(t)   =  \omega^{AB}t  + \Delta \varphi(t)$, where 
\begin{equation}
\Delta \varphi(t) =  \frac{1}{2}  (\omega^{AA} -  \omega^{BB})(\Delta t^{AA} - \Delta t^{BB}). \label{varphi}
\end{equation}
Here, $\omega^{AB} = (E_0^{AB} - E_1^{AB})/\hbar$,  $\omega^{AA} = (E_0^{AA} - E_1^{AA})/\hbar$ and $\omega^{BB} = (E_0^{BB} - E_1^{BB})/\hbar$  are the differences between the energies of the high-field stationary states in the corresponding triplet-pair configuration, and each one of them can be negative  as well as positive.

The average random-walk dwell-time in any triplet-pair configuration is  $\tau/2$, where $\tau$ is the average hopping time of individual triplet excitons between inequivalent sites. It follows that a random walk of duration $t$ averages $N = 2 t/\tau$ steps, with half the time spent in either a $AA$ or $BB$ configuration. The variance of the term $(\Delta t^{AA}~-~\Delta t^{BB})$ in (2) is then $\tau t/2$, and the variance of $\Delta \varphi(t)$ becomes
\begin{equation}
	 \langle \Delta \varphi(t)^2 \rangle =   \frac{1}{8}  ( \omega^{AA} -   \omega^{BB})^2 \tau t . \label{TIDphaseshift}
\end{equation}
It follows that the probability $|\braket{\psi(0)}{\psi(t)}|^2$ will be modulated by a function of the kind $1+ a f(t)$, where $a>0$ is a small amplitude and $f(t)$ is found by  integrating phase-shifted sinusoidal oscillations weighted by the probability of each phase, described by a normal distribution with the above variance, as in
\begin{equation}
f(t) = \int_{- \infty}^{\infty} \cos( \omega^{AB} t + \varphi) \frac{e^{-{\varphi^2}/(2 \langle \Delta \varphi(t)^2 \rangle)}}{\scriptstyle \sqrt{ \langle \Delta \varphi(t)^2 \rangle}} d\varphi  . \label{GaussianPhaseInt}
\end{equation}
Interestingly, this integral is analogous to that obtained under the gaussian phase approximation  for the spin echo signal in nuclear magnetic resonance \cite{Douglass58,Ziener18}. Integration leads  to $f(t) = \exp(-k_{TID} t) \cos \omega^{AB}t$, where
\begin{equation}
k_{TID} =   \frac{\tau}{16} (\omega^{AA} -   \omega^{BB})^2 \label{TIDrate}
\end{equation}   
is the TID-induced decay rate of the quantum beats envelope. We confirmed that this prediction matches the results obtained in a Monte Carlo simulation of the same process.  Eq.~\ref{TIDrate} implies a rule of thumb for the hopping time for which TID halves  the quantum beat amplitude after one oscillation period: in terms of frequencies, $\tau_{1/2} =  0.286 f/\Delta f^2 \approx  2 f/(7 \Delta f^2)$, where $f=\omega_{AB}/(2 \pi)$ and  $\Delta f=f_{AA}-f_{BB} = (\omega^{AA} -   \omega^{BB})/(2 \pi)$.

For  hopping times $\tau \ll \tau_{1/2}$, the triplet-exciton Hamiltonian becomes effectively the average of those of the inequivalent sites,  a situation similar to tetracene \cite{Sternlicht61,Yarmus72}, with persistent quantum beats both at zero field and for any direction of applied magnetic fields. The fact that shortening  of the hopping time  leads to  more persistent beats with a well-defined frequency is similar to  motional line narrowing  in nuclear magnetic resonance \cite{Bloembergen48}. 

As the hopping time increases past $\tau_{1/2}$, a large $k_{TID}$ will destroy any quantum beats. Then, at even larger hopping times $\tau$ such that $\tau \gg 2 \pi/|\omega^{AA}|$ and $ \tau \gg  2 \pi/|\omega^{BB}|$ the triplet pairs would remain confined to the triplet-pair configuration in which they are born, leading to  high-field quantum beats that last for a time of the order of the hopping time and contain the frequencies of each configuration.

\begin{figure}
\includegraphics[width=\columnwidth]{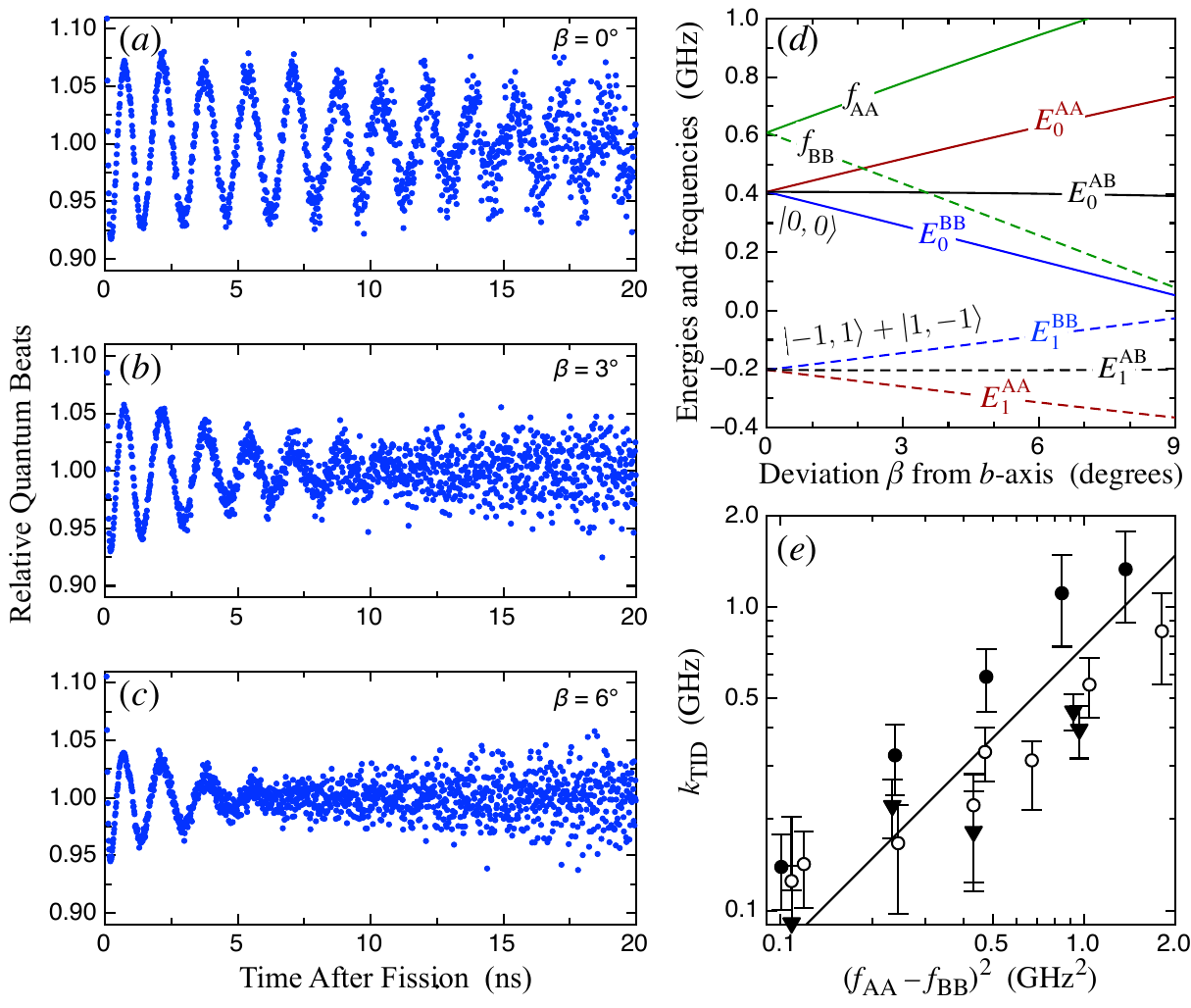}
\caption{($a$-$c$): Quantum beats  in the ratio between fluorescence and its non-oscillating trendline, as a 0.3 T magnetic field  is rotated by an angle $\beta$ from the rubrene $b$-axis in the plane bisecting the $ac$-axes. ($d$): Energies and frequencies associated with different triplet-pair configurations  as they diverge with magnetic field angle $\beta$ (the energies  near 0.4 GHz at zero-field belong to the $\ket{0,0}$ state, those near  $-0.2$ GHz belong to the $\ket{-1,1}+\ket{1,-1}$ state).
($e$) Experimental values for $k_{TID}$ of Eq.~(\ref{TIDrate}) as a function of the frequency difference between triplet-pair configurations for different magnetic field orientations near the main symmetry axes (open circles for the low frequency beats near the $b$-axis, filled circles for the intermediate frequencies near the $c$ axis, and filled triangles for the highest frequency beats near the $a$-axis). The solid line is the prediction of Eq.~(\ref{TIDrate}) for $\tau = 300 {\rm \ ps}$.} \label{beats}
\end{figure}

The result of Eq.~\ref{TIDrate} offers the opportunity to directly quantify the TID effect experimentally, and also determine  the hopping time between inequivalent sites  for  the case of rubrene, which corresponds to the hopping time $\tau_a$ in the $a$ direction (see Fig.~\ref{RubreneXstalViews}). 
We use time-correlated single photon counting to measure the time-evolution of the delayed fluorescence emitted after impulsive excitation of a singlet exciton population under low excitation conditions where the only detected photons arise from geminate fusion \cite{Wolf18}. Under these circumstances the probability of detecting a photon is proportional to $N_{TT}(t) p_\gamma(t) [1 + a f(t)]$, where  $N_{TT}(t)$ is the total number of triplet pairs, and $p_\gamma(t)$ is the probability of a re-encounter \cite{Wolf21}. We then extract the quantum beat signal, proportional to $f(t)$, by fitting the transient fluorescence data with a non oscillating trendline corresponding to $N_{TT}(t) p_\gamma(t)$ and dividing the experimental data by it. This is the same procedure followed in Ref.~\onlinecite{Wolf18}, and it is worth stressing the importance of taking the ratio between signal and non-oscillating trendline, not the difference as has been usually done in the literature. 

Fig.~\ref{beats}$a$-$c$ shows the quantum beats we extracted from the fluorescence signal   as the magnetic field is tilted   away from one of the symmetric directions in rubrene (the $b$-axis). While we observe quantum beats to persist at least for 50 ns (before they disappear into the noise)  in all cases when the magnetic field is perfectly aligned with the symmetry axes between inequivalent molecules,  varying the magnetic field direction  away from each  symmetry axis causes  $k_{TID}$, from Eq.~(\ref{TIDrate}), to increase from zero,   leading to faster and faster exponential decay rates of the quantum beat amplitude.   We  measured this effect starting from different symmetric magnetic field directions and in various geometries. The results are plotted in  Fig.~\ref{beats}$e$ as a function of the (angle-dependent) calculated value of  $(\omega^{AA} -   \omega^{BB})^2$. The plot indeed shows  the quadratic dependence  predicted  by Eq.~(\ref{TIDrate}).   Both the observed exponential decays of the quantum beat envelope and Fig.~\ref{beats}$e$   are a clear confirmation of TID and of our model for it. As shown in Fig.~\ref{beats}$e$, the experiments agree well with the prediction of  Eq.~ (\ref{TIDrate}) for a  triplet-exciton hopping time in the $a$ direction of the order of $\tau=\tau_a = 300 \pm 150$~ps.  As a comparison, the hopping time in rubrene in the  high-mobility $b$ direction can be estimated from the corresponding diffusion constant  to be less than $\sim 3$~ps  \cite{Wolf21}.

 In conclusion,  we demonstrated that transport-induced dephasing (TID)  occurs while the constituents of an entangled triplet-exciton pair hop between inequivalent sites,  and that it can be an important feature in organic molecular crystals.  
 We have also demonstrated  quantum beat measurements as a function of magnetic field orientation as a  tool that can provide information on triplet-exciton localization on inequivalent molecular sites, and their hopping rate between them.
Last but not least, we have shown that exciton hopping, inequivalent sites, and entangled triplet pairs are a fundamentally interesting condensed-matter quantum system where the decay of global coherence within an ensemble can be  tuned by the direction of the magnetic field, while the local coherence and entanglement are indefinitely maintained (for more than 50 ns in rubrene single crystals at room temperature).

\begin{acknowledgements}
Research   supported by the US Department of Energy, Office of Basic Energy Sciences, Division  of Materials Sciences and Engineering, under Award No.~DE-SC0020981.
\end{acknowledgements}


%

\end{document}